\documentclass[referee,naturemag]{sn-jnl}
\usepackage{lineno,hyperref,soul,multirow}
\usepackage[utf8]{inputenc}
\usepackage{amsmath,amssymb,color,xcolor,gensymb}
\usepackage{graphicx,caption,subcaption} 
\usepackage{amsfonts,amsthm,amsthm} 
\usepackage{graphicx,float} 
\usepackage{upgreek}
\modulolinenumbers[5]
\usepackage{geometry}
\begin{document}

\title[Chemically driven defect formation]{
Metastable defect phase diagrams as a tool to describe chemically driven defect formation: Application to planar defects}

\author*[1]{\fnm{Ali} \sur{Tehranchi}}\email{tehranchi@mpie.de}

\author*[1]{\fnm{Siyuan } \sur{Zhang}}\email{ siyuan.zhang@mpie.de}

\author[1]{\fnm{Ali} \sur{Zendegani}}
\author[1]{\fnm{Christina} \sur{Scheu}}\email{c.scheu@mpie.de}

\author[1,2]{\fnm{Tilmann} \sur{Hickel}}\email{hickel@mpie.de}
\author[1]{\fnm{J\"org} \sur{Neugebauer}}\email{neugebauer@mpie.de}
\affil[1]{ \orgname{Max-Planck-Institut f\"ur Eisenforschung GmbH}, \orgaddress{\city{D\"usseldorf}, \postcode{D-40237}, \country{Germany}}}
\affil[2]{ \orgname{BAM Federal Institute for Materials Research and Testing}, \orgaddress{\city{Berlin}, \postcode{D-12489}, \country{Germany}}}

\abstract{
Thermodynamic bulk phase diagrams have become the roadmap used by researchers to identify alloy compositions and process conditions that result in novel materials with tailored microstructures. Recent experimental studies show that changes in the alloy composition can drive not only transitions in the bulk phases present in a material, but also in the concentration and type of defects they contain. Defect phase diagrams in combination with density functional theory provide a natural route to study these chemically driven defects. Our results show, however, that direct application of thermodynamic approaches can fail to reproduce the experimentally observed defect formation. Therefore, we extend the concept to metastable defect phase diagrams to account for kinetic limitations that prevent the system from reaching equilibrium. We successfully applied this concept to explain the formation of large  concentrations of planar defects in supersaturated Fe-Nb solid solutions and to identify in a joint study with experiments conditions in Mg-Al-Ca alloys for defect phase occurrence. The concept offers new avenues for designing materials with tailored defect structures.}

\maketitle

\section{Introduction}\label{Sec:intro}
The microstructure of a material is often characterized by complex patterns resulting from the presence of several bulk phases and precipitates. Notable examples of this are bainite or pearlite steels~\cite{djaziri2016deformation}. In order to design materials with these intricate microstructures, it is necessary to accurately predict the alloy compositions and process temperatures at which the desired phases will form. To meet this need, modern computational tools such as CALPHAD (Calculation of Phase Diagrams) have been developed~\cite{saunders1998calphad,sundman2007computational,liu2009first}. Recent advances in experimental techniques have enabled researchers to study materials at the atomic scale, providing detailed information about the chemical and structural nature of bulk phases as well as extended defects such as grain boundaries and dislocations~\cite{zhang2022atomistic,meiners2020observations,cantwell2014grain,rhode2013mg}. These high-resolution studies have allowed for the characterization of defects in terms of their local composition, size, and structure as a function of alloy composition and temperature, leading to the development of defect or complexion phase diagrams. 

It is generally assumed that the bulk phase diagram, which represents the vast majority of the atoms in a material, accurately predicts the experimentally observed bulk phases. However, a recent study on Fe-Nb found that in a region of the phase diagram where the Fe\textsubscript{2}Nb-C14 Laves phase and the Fe\textsubscript{7}Nb\textsubscript{6}-$\mu$-phase were expected to coexist, solely the C14 phase with a large number of Nb-rich planar defects were observed instead~\cite{vslapakova2020atomic}. Thus, the supersaturated Nb-rich Fe\textsubscript{2}Nb phase formed after cooling did not lead to the formation of $\mu$-phase precipitates indicated by the binary Fe-Nb bulk phase diagram. Instead, the excess Nb resulted in the formation of planar defects and a microstructure that differed from what would be expected based on the bulk phase diagram. This is highly significant because the microstructure of a material can strongly influence its mechanical or corrosion properties~\cite{korte2022defect}. In order to effectively use this mechanism for a wide range of materials systems, it is essential to have a computational approach that is able to treat both bulk and chemically driven defect phases on the same level.

In this letter we first outline a general approach that combines density functional theory (DFT) calculations with grand-canonical thermodynamic concepts. As will be shown for the Fe-Nb system, assuming thermodynamic equilibrium fails to reproduce the experimentally observed deviation from the bulk phase diagram. We therefore explicitly include in this approach the presence of kinetic barriers and show that they are crucial to faithfully reproduce experimental observations. Having demonstrated the validity of our approach for the Fe-Nb system, we employ it for a model Mg-Al-Ca alloy to identify conditions for defect phase formation. To further validate this approach, we use the identified alloy and process conditions to cast the corresponding Mg alloy and perform high resolution transmission electron microscopy measurements to show the predicted defect formation.

\section{Metastable defect phase diagrams}\label{Sec:concept}
In our proposed approach, we will use the chemical potential of the solute species as a key quantity to understand and quantify the reaction that converts excess solutes into either bulk precipitates (a secondary bulk phase forms according to the traditional bulk phase diagram) or into a solute-rich defect phase. In order to compare the two qualitatively very different reactions on equal footing, the chemical potential of the resulting (reactant) phase has to include both the energy required to form the new lattice/defect structure and the energy gained during its formation in this phase (i.e., the segregation energy). We note that this choice is very different from the often-considered case of conventional segregation, where defects are already present. In this case, the energy cost for defect formation is not needed resulting in much lower chemical potentials. However, for the case we are interested in -- solute-induced defect formation -- this contribution is crucial~\cite{weissmuller1993alloy,kirchheim2007reducing}.  In fact, it is exactly the competition between endothermic defect formation and exothermic segregation that controls solute-driven defect formation \cite{nazarov2010first,nazarov2014ab}. This approach, which includes both contributions, is also consistent with the treatment of bulk phases.

\begin{figure}[ht]
    \centering
    \includegraphics[width=\textwidth]{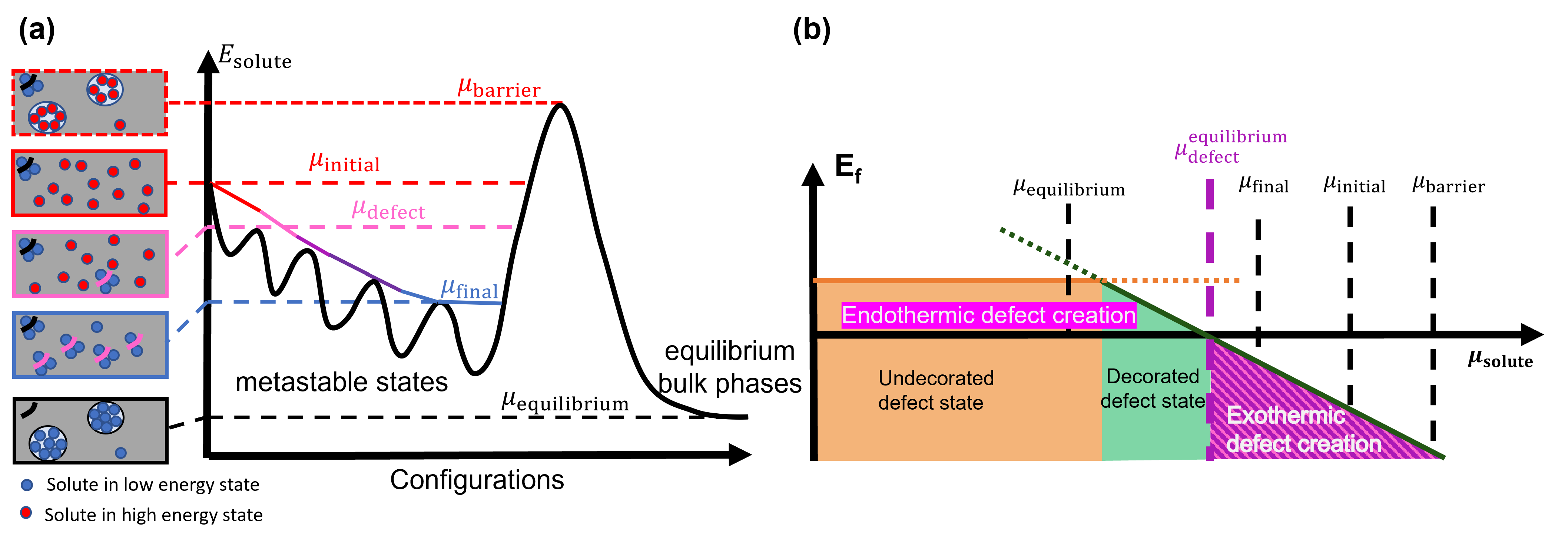}
    \caption{Schematic representation of the equilibration process (a) and the metastable defect phase diagram (b) as discussed in the text. Non-equilibrium (excess) solutes are marked by red dots. Energetically favorable solute configurations, e.g., in the bulk-like precipitates or at defects are shown as blue dots. The chemical potentials related to the various configurations sketched in the gray boxes are given by the dashed horizontal lines in (a) and vertical lines in (b).}
    \label{Fig:nucleation}
\end{figure}

The change of the total energy of the system per excess solute, $E_{\rm solute}$, versus different configurations is sketched schematically in Fig.~\ref{Fig:nucleation}a. At casting conditions, i.e., close to the melting temperature, the system is in a single-phase solute solution containing defects. When the system is quenched to a lower temperature, where the solute concentration is above the solubility limit, at this temperature, the chemical potential of the initial  supersaturated phase is $\mu_{\rm initial}$. This phase is thermodynamically unstable against the formation of precipitates.  
The resulting precipitates provide an energetically favorable configuration for the excess solutes. The chemical potential for the precipitate is $\mu_{\rm equilibrium}$ and well below the initial chemical potential. The (positive) difference between the two potentials $\mu_{\rm initial} - \mu_{\rm equilibrium}$ drives the partitioning and precipitate formation. 

As outlined above and implied by the experimental observations an alternative way to accommodate the excess solutes in the initial solid solution phase is the formation of defects that provide energetically more favorable sites than the original matrix. The chemical potential to form and occupy the defect is $\mu_{\rm defect}^{\rm equilibrium}$. 
In the absence of any reaction barriers, the reaction with the lowest chemical potential in the final state would dominate. In many cases, $\mu_{\rm equilibrium} < \mu^{\rm equilibrium}_{\rm defect}$. As a consequence, solute-driven exothermic defect formation should not occur. Since this is in obvious contradiction to the experimentally observed presence of these defects, the assumption of barrier-less transitions does not hold. We therefore include in the following discussion also kinetics. For the formation of precipitates, the nucleation barrier has to be overcome. This barrier can be rather large. The chemical potential for a solute in such a barrier configuration is $\mu_{\rm barrier}$. The presence of this barrier has an important consequence: As schematically shown in Fig.~\ref{Fig:nucleation}a the presence of this barrier makes the thermodynamically unstable (against the precipitate bulk phase) defects metastable. Thus, as long as the chemical potential is below the barrier configuration, the dominant reaction will be defect rather than precipitate formation. Since metastability plays a key role in enabling and driving this unconventional solute-driven defect formation that even outperforms bulk formation we call the formalism metastable defect phase diagram (MDPD).  

Based on the above discussion, the formalism can be schematically summarized in Fig.~\ref{Fig:nucleation}b. The central quantities are the defect formation energies as a function of the solute chemical potential. This representation has been originally developed to describe point defects in semiconductors~\cite{zhang1991chemical} and has been later extended to surface phase diagrams~\cite{freysoldt2014first}. An important advantage of this approach is that it can be readily connected to ab initio calculations since it requires only DFT computed energies and the number of atoms for each species as input
\begin{align}
    E_{\rm f}^{\rm defect}=E_{\rm tot}^{\rm defect}-E_{\rm tot}^{\rm bulk}-n_{\rm solute}\mu_{\rm solute}\label{Eq:nb},
\end{align}
where $E_{\rm tot}^{\rm defect}$, $E_{\rm tot}^{\rm bulk}$, $n_{\rm solute}$, and $\mu_{\rm solute}$ are the total energy of the supercells containing defects and defect-free bulk, the number of excess solute atoms segregated to the defects, and the chemical potential of solute atoms, respectively.  Fig.~\ref{Fig:nucleation}b shows exemplarily the formation energy of two defect states (orange and green solid lines). The different slopes reflect the different coverages of solute atoms at the defect. We use coverage, as commonly used in surface science, instead of concentration to reflect that the excess solute concentration at a planar defect is referenced to an area and not to the bulk volume.      

The mechanism that leads to the experimentally observed formation of solute-rich defects can be described as follows (see Fig.~\ref{Fig:nucleation}b): If the initial chemical potential of the supersaturated solid solution is less than the hypothetical chemical potential solute atoms would have in the nucleation barrier $\mu_{\rm barrier}$, the formation of the  bulk precipitates ($\mu_{\rm equilibrium}$), though thermodynamically preferred, is kinetically prohibited. 
An alternative reaction path to lower the energy of the excess solutes is the formation of extended defects, which provide for the solutes binding sites that are energetically much more favorable than the substitutional sites in the solid solution. The final chemical potential is found in most cases above the bulk equilibrium one ($\mu_{\rm defect}^{\rm equilibrium} > \mu_{\rm equilibrium}$), i.e., the defect phases are thermodynamically metastable. However, since barriers to form the lower dimensional defects are typically lower than the bulk nucleation ones, defect formation may take precedence over bulk nucleation.

 The small offset between $\mu_{\rm final}$ and $\mu_{\rm defect}^{\rm equilibrium}$ shown in Fig.~\ref{Fig:nucleation}b is given by the defect nucleation barrier. It arises, since a defect in contrast to a bulk precipitate can only host a finite number of solutes. Thus, once a defect becomes completely filled a new one needs to be created. Once the chemical potential gets below this barrier, the formation of defects ceases and a metastable equilibrium is reached. 
 
\section{Competition between bulk precipitates and Nb-rich planar defects}\label{Sec:SFNB}
To demonstrate the applicability and the predictive power of this approach in connection with modern DFT calculation we first consider the Fe-Nb system.  
In this material system, the presence of the C14 Laves phase has been a topic of many theoretical and experimental studies~\cite{liu2012ab,voss2011phase,stein2015phase}. The reason is that Laves phases in steels are often brittle at ambient temperature, but their high melting temperature makes them very stable and a suitable constituent in (ferritic) steels serving at elevated temperatures~\cite{knezevic2002martensitic,falat2005mechanical,aghajani2009effect}.  

\begin{figure}[ht]
     \centering
         \includegraphics[width=1.0\textwidth]{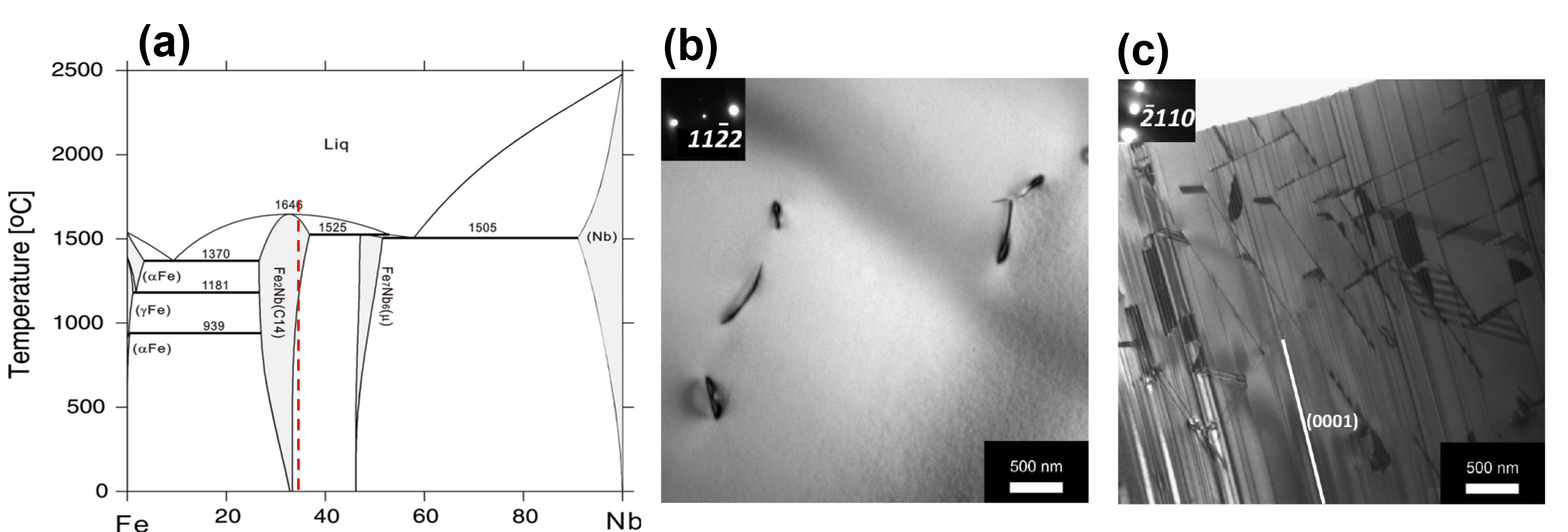}
        \caption{(a) The Fe-Nb phase diagram, reprinted with permission from Ref.~\cite{he2017thermodynamic}. Bright field-TEM images of (b) a stoichiometric alloy with 33 at.\% Nb and (c) a Nb-rich alloy with 35 at.\%, reprinted with permission from Ref.~\cite{vslapakova2020atomic}. Elsevier 2020.}
        \label{Fig:FeNb}
\end{figure}

 Vo{\ss} et al.~\cite{voss2011phase} determined experimentally the C14-NbFe$_2$ Laves phase on both sides of its stoichiometric 33.3 at.\% Nb composition, i.e., between 25.1 at.\% and 37.6 at.\% Nb. 
According to the Fe-Nb system phase diagram~\cite{he2017thermodynamic} as shown in Fig.~\ref{Fig:FeNb}a, an alloy with an overall composition between 35 at.\% and 37.5 at.\% Nb is in the single-phase region at high temperature, yet at lower temperatures it enters the two-phase region of C14+$\upmu$ phases (see vertical dashed line in Fig.~\ref{Fig:FeNb}a). Therefore, when cooling down from high temperatures to room temperature bulk thermodynamics predicts the precipitation of the Nb-rich $\upmu$ phase (Nb$_6$Fe$_7$). 

Šlapáková et al.~\cite{vslapakova2020atomic} performed a transmission electron microscopy (TEM) study on single-phase stoichiometric (33 at.\% Nb) as well as Nb-rich  (35 at.\% Nb) NbFe$_2$ Laves phase specimens to characterize their microstructures.  In their work, representative microstructures of the stoichiometric and the Nb-rich alloy were shown by bright-field TEM images. The stoichiometric sample only reveals the presence of isolated dislocations with a comparatively low defect density (see Fig.~\ref{Fig:FeNb}b). However, for Nb-rich alloys bright-field TEM did not detect precipitates but a  large increase compared to the stoichiometric samples of extended and confined planar defects (Fig.~\ref{Fig:FeNb}c).  An energy dispersive X-ray spectroscopy (EDS) analysis performed in a scanning TEM (STEM) showed that the Nb concentration at these planar defects is largely increased compared to its bulk concentration. These experimental observations are in obvious contrast to the above-mentioned prediction of the bulk phase diagram,  according to which the 2 at.\% excess Nb content should result in the formation of $\upmu$-precipitates.

\begin{figure}[ht]
    \centering
    \includegraphics[width=\textwidth]{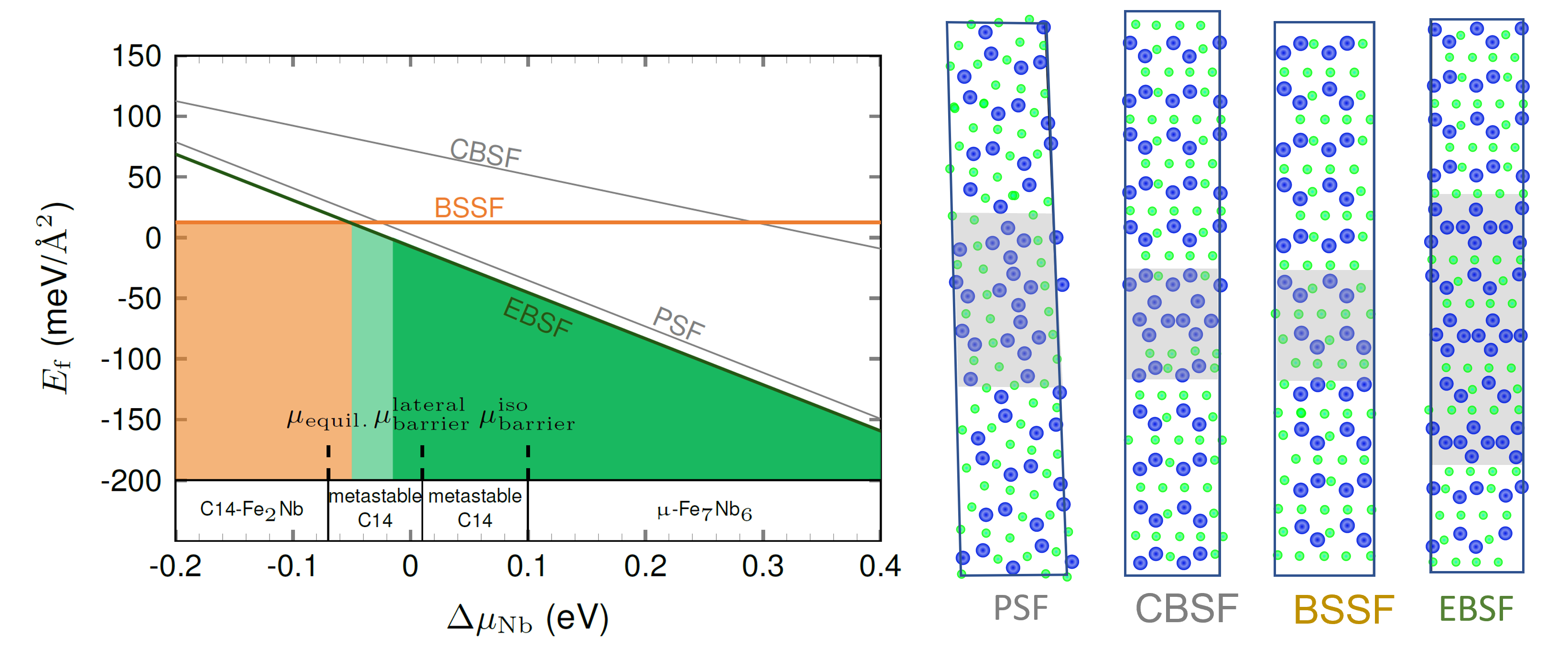}
    \caption{The energies of the formation of structures with and without defects are plotted as a function of the Nb chemical potential $\mu_{\rm Nb}$ according to Eq.~\eqref{Eq:nb}. The thermodynamically (meta)stable bulk phase at a given chemical potential domain is given in the white box at the bottom. The color coding labels the nature of defect formation: In the orange region the undecorated defect state is dominant. In the light green and green regions, the decorated defect state is endothermic and exothermic, respectively.  The atomic structures of extended basal stacking fault (EBSF), pyramidal stacking fault (PSF), confined basal stacking fault (CBSF), and basal synchroshear-formed stacking fault (BSSF), are sketched in the right panel.  The blue, and green spheres denote Nb, and Fe atoms, respectively.  The planar defects are highlighted by the gray regions. }
    \label{fig:DPD_nb}
\end{figure}

Fig.~\ref{fig:DPD_nb} shows the calculated energies of formation for the planar defects as a function of the chemical potential of Nb, $\Delta\mu_{\rm Nb}$.  Here $\Delta\mu_{\rm Nb}=\mu_{\rm Nb}-\mu_{\rm Nb}^0$ where $\mu_{\rm Nb}^0$ is the chemical potential of Nb atoms in bulk Nb. 
The slope of the line describing the formation energy of a given defect  is proportional to the amount of the excess Nb atoms of this defect phase. The excess is defined with respect to the composition of the C14-NbFe$_2$ Laves phase. Consequently, the defect with zero excess Nb atoms (BSSF), i.e. the defect-free C14-NbFe$_2$ Laves phase, has zero slope.

In the region of chemical potentials with $\Delta\mu_{\rm Nb} \le \mu_{\rm equil.}=-0.07$ eV the energetically most stable bulk phase is the C14-NbFe$_2$ Laves phase.  If the Nb chemical potential exceeds this potential $\mu_{\rm equil.}$, the $\upmu$ phase becomes the most stable phase. 
The positive defect formation energies in Fig.~\ref{fig:DPD_nb} for $\Delta\mu_{\rm Nb} \le -0.07$ eV imply that none of the experimentally detected defect structures is thermodynamically stable: From a thermodynamic perspective a phase separation into the $\upmu$ phase and C14 Laves phase should occur. 
The defects are even unstable against the formation of undecorated basal synchroshear-formed stacking faults (BSSF).
We can therefore conclude that the defect phases observed experimentally cannot be explained when assuming that the system has reached thermodynamic equilibrium. As a consequence, the bulk phase diagram alone is unable to tell us into which phases the excess solutes partition or what are the actual (i.e., non-equilibrium) chemical potentials in the sample.

As outlined in the discussion of Fig.~\ref{Fig:nucleation}a the existence of a large nucleation barrier prevents the formation of the bulk precipitates, resulting in a supersaturated solid solution with a chemical potential for the solutes well above that at which the bulk precipitates could form thermodynamically. An upper limit of the solute chemical potential is given by the chemical potential $\mu_{\rm barrier}$ the solutes have in the precipitate phase. Once the potential reaches this value precipitates form spontaneously without having to overcome a barrier. The chemical potential $\mu_{\rm barrier}$ is thus a key quantity to extend defect phase diagrams to supersaturated conditions not covered by bulk phase diagrams.

In the following we will therefore introduce two models. Together with DFT simulations they will allow us to estimate values for $\mu_{\rm barrier}$. For substitutional solutes as considered here, the initial nucleus will be a coherent agglomeration of solutes in a sphere, with a solute concentration close or equal to that of the bulk precipitate. Since we are interested in a lower limit of the barrier chemical potential, we can safely neglect the interface energy between nucleus and matrix. The barrier configuration is then described by the precipitate bulk system, except that the lattice constant is not that of the plastically relaxed precipitate, which would have the chemical potential $\mu_{\rm equillibrium}$ of the bulk phase. Rather, the lattice constant is dictated by the coherency to the bulk resulting in significantly higher chemical potentials. 

For homogeneous nucleation coherence extends in all three spatial directions. The chemical potential for this barrier (nucleation) configuration is $\mu_{\rm barrier}^{\rm iso}=\mu_{\rm equil.}+ (E_{\rm sol}^{\rm DFT}(a_{\rm coh}) - E_{\rm sol}^{\rm DFT}(a_{\rm eq}))/n_{\rm sol}$ and can be directly obtained via DFT calculations. $E_{\rm sol}^{\rm DFT}(a)$ is the total energy of the precipitate phase at the coherency lattice constant $a_{\rm coh}$ and the equilibrium lattice constant $a_{\rm eq}$. $n_{\rm sol}$ is the number of solute atoms in the corresponding supercell.

In case of nucleation at basal defects, isotropy is broken.  The coherency constrain may thus involve only the two lateral lattice constants, with the lattice constant normal to the basal plane being allowed to relax. The corresponding chemical potential is $\mu_{\rm barrier}^{\rm lateral}= \mu_{\rm equil.}+(E_{\rm sol}^{\rm DFT}(a^{\parallel}_{\rm coh}, a^{\perp}_{\rm eq}) - E_{\rm sol}^{\rm DFT}(a^{\parallel}_{\rm eq}, a^{\perp}_{\rm eq}))/n_{\rm sol}$. The values of $\mu_{\rm barrier}^{\rm lateral}$ and  $\mu_{\rm barrier}^{\rm iso}$ are 0.01\,eV and 0.1\,eV, respectively. These values are substantially larger than the maximum Nb chemical potential that can be reached in the C14 phase in thermodynamic equilibrium. These values are highlighted in the metastable phase diagram which is displayed in the white box at the bottom of Fig.~\ref{fig:DPD_nb}. 

As shown in Fig.~\ref{fig:DPD_nb} at such high chemical potentials defect states become thermodynamically favored that were absent in the equilibrium case. Specifically, a chemically ordered extended basal stacking fault EBSF becomes stable. This defect is for intermediate values of the Nb chemical potential endothermic (light green), i.e., this defect phase will form only at existing basal stacking faults. At even higher Nb chemical potentials defect formation becomes exothermic: The system will create such defects spontaneously. Since the creation of these defects takes solute atoms out of the host matrix the Nb chemical potential will decrease until defect formation becomes endothermic. At this point the driving force for spontaneous defect formation vanishes and the formation of further defects stops. 

The mechanism outlined here is to a large part very similar to lowering the solute chemical potential via bulk-like precipitates. 
The main differences are that precipitate
formation is fully covered by the bulk phase diagram, i.e., an oversaturated solid solution partitions into precipitates and the host matrix. Like precipitates, exothermic defects coexist with the host matrix and remove large amounts of solutes from the surrounding host matrix. In this scenario the defects can no longer be regarded as low-dimensional objects having an infinitely small volume fraction but control like bulk precipitates phase equilibration and coexistence. 

The MDPD, i.e., the defect phase diagram in combination with the upper limit for the chemical potential due to the nucleation barrier shown in Fig.~\ref{fig:DPD_nb}, provides a direct quantitative approach to describe phase coexistence between defects and bulk. It is a natural extension of both defect phase and bulk phase diagrams and
explains the experimentally observed formation of large concentrations of Nb rich planar defects in the C14 phase. The theoretical approach of  MDPDs outlined here is general and not restricted to the specific Fe-Nb alloy system considered in this section. In the next section we will use it to identify conditions where exothermic defect formation can be induced in Mg-based alloys.

\section{Using Metastable defect phase diagrams to induce planar defects in ternary Mg-Al-Ca alloys}\label{Sec:SFMG}
To demonstrate the exciting opportunities opened by ab initio computed MDPDs in materials design we will use them in the following to identify suitable process conditions where chemically driven defect formation takes precedence over bulk precipitates. As prototype system we will consider Mg-alloys, which are  
highly attractive for, e.g., automotive applications due to their low cost and weight~\cite{zubair2021co,guenole2021exploring}. We will focus on Mg alloyed with Al and Ca, since adding small amounts of these elements largely improves the ductility of Mg \cite{sandlobes2017rare,pei2015rapid}. Going above the solubility limit, Al$_{2-x}$CaMg$_x$ Laves phases can be formed, where the Al- ($x$=0) and Mg-rich ($x$=1) stable Laves phases are Al$_2$Ca-C15 and CaMg$_2$-C14, respectively.

\begin{figure}[ht]
    \centering
    \includegraphics[width=\textwidth]{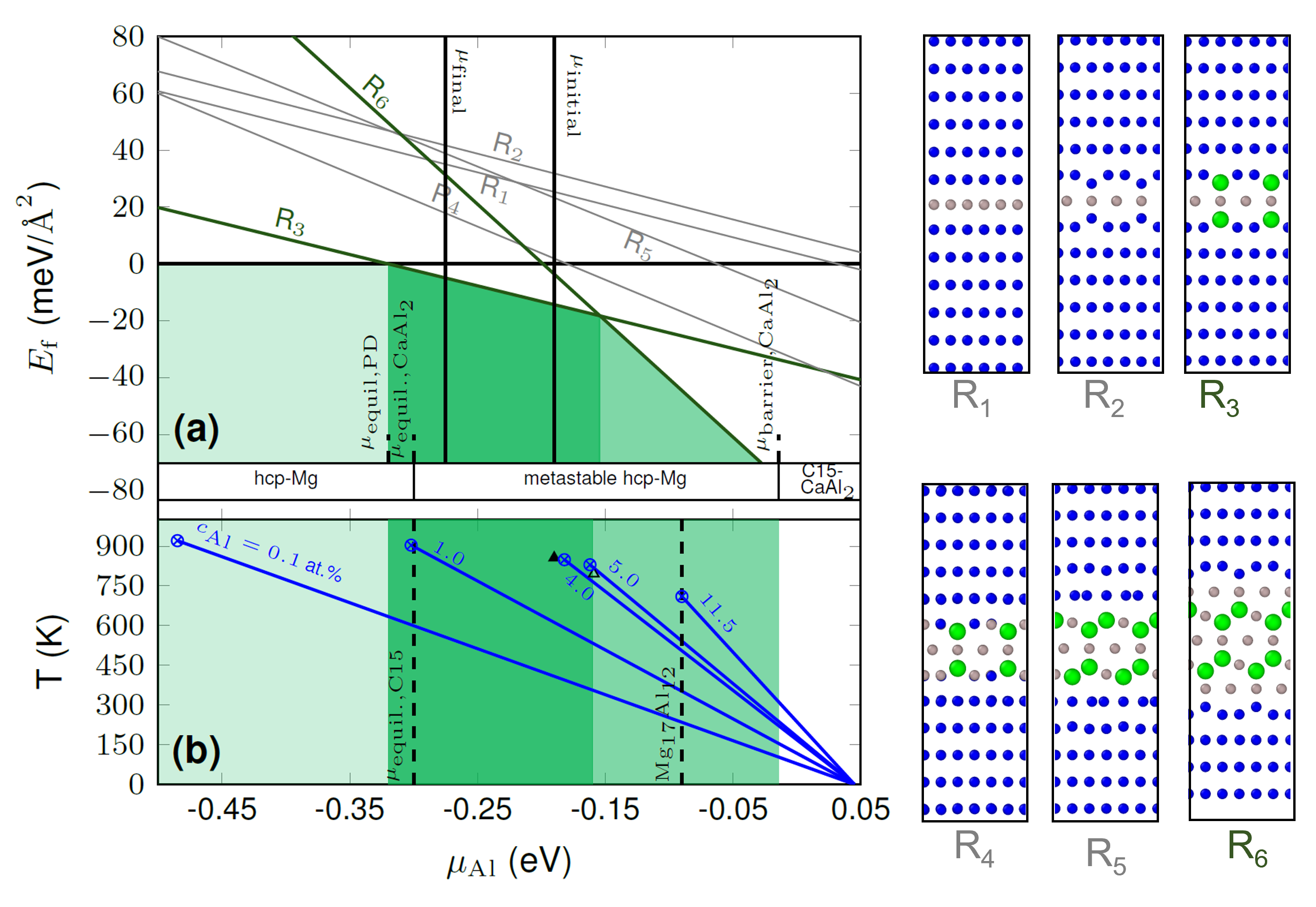}
    \caption{ (a) Formation energies (Eq.~\eqref{Eq:nb}) as a function of the Al chemical potential for various planar structures with and without defects. The thermodynamically (meta)stable bulk phase for a given chemical potential are given in the bottom white box.  The dark green region denotes the exothermic formation of the R$_3$ defect.  The light green regions at left and right of it denote the defect-free and the exothermic R$_6$ formation regions, respectively.
    (b) Variation of the temperature of solid solution versus chemical potential for various concentrations of Al.  The melting temperature for each alloy is highlighted by the $\otimes$ symbol.  The experimental conditions of Ref.~\cite{Kashiwase2019} and the present work are highlighted by hollow and filled triangles, respectively.  The atomic structures of the different planar defects are given in the right panel.  The blue, green and gray spheres denote Mg, Ca, and Al atoms, respectively. }
    \label{Fig:chemicalLand}
\end{figure}

Platelets of C15-like atomic arrangements are known to form along the Mg basal plane (0002)\cite{Suzuki2008, zubair2022laves}. The small lattice mismatch between the in-plane directions of Al$_2$Ca(111) and Mg(0002), makes the C15 phase also an interesting candidate for forming planar defect structures within the Mg matrix. 
Experiments of Kashiwase et al.~\cite{Kashiwase2019,Kashiwase2020} using high-resolution TEM measurements show that the platelets of C15 can be thinner than the bulk unit cell of C15.  
However, the atomic structure and chemical distribution of elements of such C15-like planar defects remained unclear.
To explore whether such defects consisting of a few atomic layers can be exothermically formed we study their formation energy. Specifically, we consider a series of defects starting from a single Al layer and sequentially adding further layers. The first six defect structures are shown in Fig.~\ref{Fig:chemicalLand} (right side) with the last one representing a single-layer C15 structure embedded in the Mg matrix.

Since the planar defects coexist with bulk Mg, we impose $\Delta\mu_{\rm Mg}=0$, i.e., the chemical potential of Mg atoms is the same as their chemical potential in bulk Mg. To maximize the amount of Ca in the matrix, as needed to drive the formation of these defects, we further consider Ca-rich conditions. These conditions are realized at or above the Ca solubility limit, i.e., the chemical potential of Ca is $\Delta\mu_{\rm Ca} = E_{\rm f}^{\rm C14-CaMg_2}-2\Delta\mu_{\rm Mg} =-0.378$ eV. Hence, the chemical potential of Al becomes the only free parameter. The reason we chose Al rather than Ca is that the solubility of Al in Mg is much higher than that of Ca.  

The corresponding MDPD is shown in Fig.~\ref{Fig:chemicalLand}a. In thermodynamic equilibrium, i.e., using the conventional bulk phase diagram,  the upper limit of the Al chemical potential in hcp Mg is $\Delta\mu_{\rm Al} = \mu_{\rm equil.}= -0.30$ eV.  In the thermodynamic stability domain of hcp Mg, defect formation becomes exothermic at $\Delta\mu_{\rm Al} = -0.315$ eV, which is just below the $\mu_{\rm equil.}$.     

To study whether defect formation can be  kinetically driven, we computed again the nucleation barrier. 
For the barrier structure we considered the bulk Al$_2$Ca-C15 Laves phase, but fixed its lattice constants to those of pure Mg.  The Al chemical potential of this highly strained phase 
is $\mu_{\rm barrier}= -0.014$ eV. This value, which sets another limit for the Al chemical potential, is substantially higher than the limit set by bulk equilibrium ($\mu_{\rm equil.}= -0.30$\,eV). In this metastable hcp Mg region defect formation becomes highly exothermic (see Fig.~\ref{Fig:chemicalLand}a). 

The MDPD shown in Fig.~\ref{Fig:chemicalLand}a allows us to extract suitable experimental process and alloying conditions to stimulate formation of defects. The thermodynamic window of Al chemical potentials at which the formation of the R$_3$ defect dominates the process is in the range of $-0.315~{\rm eV }\le \mu_{\rm initial}\le -0.16~{\rm eV}$. Since this window is well below the chemical potential of the barrier configuration ($\mu_{\rm barrier,C15}$) R$_3$ planar defects rather than bulk C15 precipitates are expected to form. The upper bound of this interval is given by the onset of R$_6$ formation. However, even for chemical potentials larger than $-0.16~{\rm eV}$ the formation of R$_6$ will lead to a decrease of the chemical potential and thus R$_6$ will decompose into R$_3$.

To translate the chemical potential into experimentally controllable quantities we use the fact that the relevant bulk Al concentrations are in the dilute limit. In this limit the chemical potential is given by 
    $\mu_{\rm Al}=\mu^0_{\rm Al} + k_{\rm B}T\ln{c_{\rm Al}}$
where  $\mu^0_{\rm Al}$ is the solution enthalpy and $c_{\rm Al}$ is the concentration of Al in the pure Mg matrix,  
$T$ is the temperature and $k_{\rm B}$ is the Boltzmann constant. 
We retrieve the solution enthalpy as $\mu^0_{\rm Al}=0.045$~eV and the concentration $c_{\rm Al}$ from the binary Al-Mg phase diagram~\cite{murray1982mg}. To this end, we impose the condition that the chemical potential of Al atoms in the solid solution and in the intermetallic Mg$_{17}$Al$_{12}$ phase should be equal for the concentration and temperature pairs lying on the equilibrium line of solid solution and intermetallic compound in the phase diagram. 

Fig.~\ref{Fig:chemicalLand}b contains the variation of temperature of solid solution versus chemical potential for various $c_{\rm Al}$. The upper limit for each line is set to the melting temperature of the solution.    The low concentrations such as 1.0 at. \% cannot provide enough driving force for formation of the defects. On the other hand, the chemical potential of  high concentration solid solutions e.g. 11.5 at.\%, will exceed the chemical potential needed for the domination of R$_3$.  But solid solutions with $\approx$ 4 to 5 at.\% near their melting temperature will fall in the suitable window of chemical potentials.
Thus, we predict that in samples with these concentrations, the formation of planar defects in as cast condition is possible. We note that the casts used in Kashiwase et al.~\cite{Kashiwase2019,Kashiwase2020} experiments in which the platelet C15 structures are detected fulfill the set forth conditions ($T$ = 796 K and $c_{\rm Al}$=4.55 at.\%).  The nominal composition of our cast designed to capture the planar defects is chosen to be Mg-4wt.\%Al-4wt.\%Ca with a composition of 1.1 at.\% Al and 0.2 at.\% Ca in the Mg matrix \cite{zubair2022laves}.  

As shown in Fig.~\ref{Fig:chemicalLand}a, the initial Al chemical potential of our cast $\mu_{\rm initial} = -0.19$ eV  is well in the range of chemical potentials suitable for an exothermic formation of R$_3$. By forming the planar R$_3$ defects the solute concentration decreases, resulting in a monotonous decrease of the solute chemical potential until at a  chemical potential $\mu_{\rm final}$ the chemical driving force is no longer sufficient to overcome the nucleation barrier for forming new defects. According to the Al concentration in defect-free bulk like regions measured by STEM-EDS the final solute chemical potential $\mu_{\rm final}=-0.27$\,eV. As shown in Fig.~\ref{Fig:chemicalLand} this value is slightly (0.05\,eV) above the onset of exothermic defect formation ($\mu_{\rm equilibrium,PD}$). We note that this small positive value is not the actual nucleation barrier of the defect. This is given by the number of solute atoms needed to form a critical nucleus of the defect times this value.

\begin{figure}[ht]
         \centering
         \includegraphics[width=\textwidth]{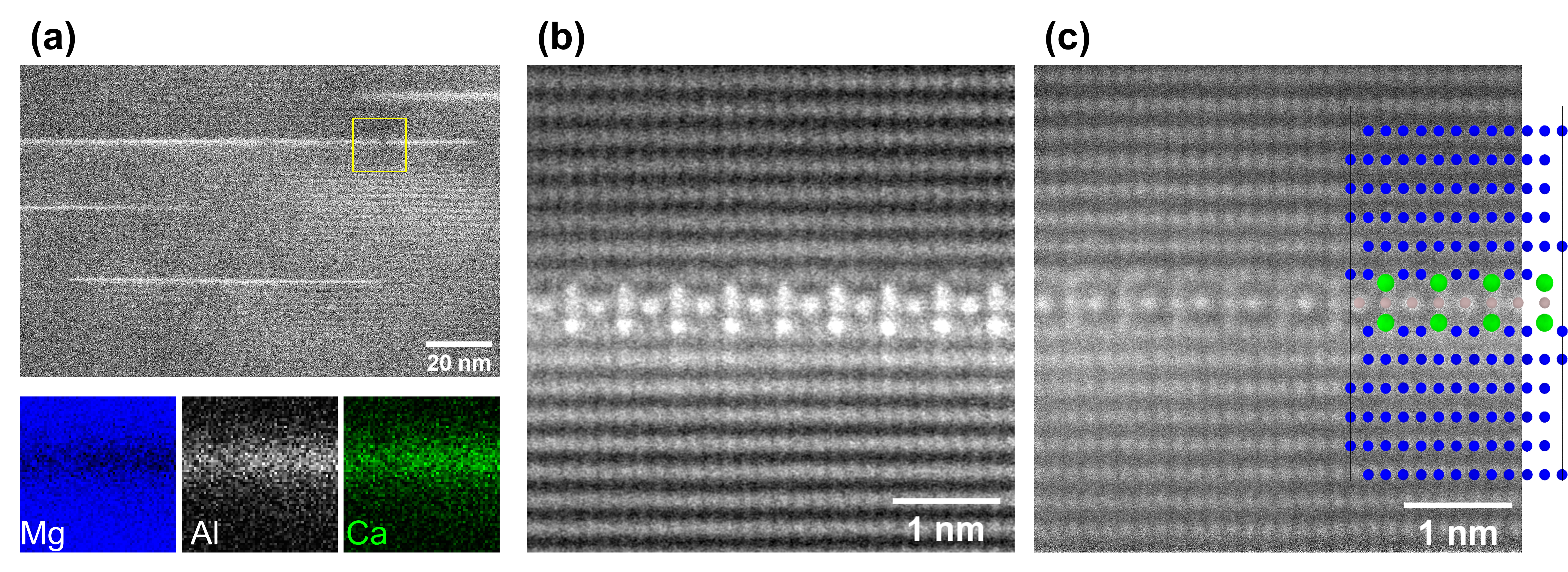}
         \caption{(a) HAADF-STEM image of planar defects in the Mg matrix of an as-cast Mg-4wt.\%Al-4wt.\%Ca alloy with corresponding EDS maps in the region highlighted by the yellow box.  The 
         EDS maps have been analized with the principal component analysis code from~\cite{Zhang2018}.  Atomic resolution HAADF-STEM images of the planar defects are acquired along Mg[1-100] (b) before and (c) after the Ca column was damaged by electron beam radiation. The corresponding atomic structures are overlaid, the blue, green, and gray spheres denote Mg, Ca, and Al atoms.}
         \label{Fig:exp}
     \end{figure}

 We imaged the planar defects in the Mg matrix in the as cast condition by atomic resolution STEM using the high angle annular dark field (HAADF) detector to provide Z(atomic number) contrast. As shown in Fig.~\ref{Fig:exp}c the R$_3$ atomic structure matches perfectly the defect observed in our HAADF-STEM images (Fig.~\ref{Fig:exp}b).  As shown in Fig.~\ref{Fig:exp}a, the planar defects are closely spaced by a few tens of nm. Their brighter contrast is consistent with the enrichment in the heavier elements Al and Ca, which are also confirmed by the EDS maps.
Atomically resolved STEM images (Fig.~\ref{Fig:exp}b,c) further reveal the orientation relationship between the C15-like planar defect and the Mg matrix: Al$_2$Ca(111) // Mg(0002), and Al$_2$Ca[11-20] // Mg[1-100]. Note that this orientation relationship has also been observed between bulk-phase C15 and Mg.\cite{Suzuki2008,zubair2022laves}
Nevertheless, it is clearly shown in Fig.~\ref{Fig:exp}b that only a fraction of a C15 unit cell is sandwiched between the Mg(0002) planes. 
The C15-like planar defect is symmetric with respect to the middle plane with alternating brighter and weaker contrast, corresponding to atomic columns with double Al and single Al occupancy. The next plane has bright contrast above the single Al occupancy column, corresponding to the Ca columns. The plane further above shows similar contrast to the Mg columns in the matrix and is believed to be either Mg or Al, which has a similar Z contrast.

It is important to note that the Ca columns in the structure are prone to electron beam damage. As shown in Fig.~\ref{Fig:exp}c, the bright contrast in the Ca column positions disappeared after prolonged image in the area. This could have hampered the structural analysis of this planar defect in earlier works. Our low electron dose images clearly identified the position of the Ca columns.

The above discussion demonstrates the power of the MDPD concept in predicting formation and structure of defects under non-equilibrium conditions as well as designing experiments to activate certain defects. It faithfully reproduces the various experimentally available observables (HR-TEM defect structures, concentrations) and allows for a quantitative description of the reaction paths qualitatively sketched in Fig.~\ref{Fig:nucleation}.   

\section{Conclusions}\label{Sec:sum}
Inspired by experimental observations showing that supersaturated alloys partition into defects rather than bulk-like precipitates, contrary to what would be expected based on their bulk thermodynamic phase diagrams, we have introduced the concept of metastable defect phase diagrams (MDPD). MDPDs provide an intuitive connection between defect phase diagrams, equilibrium bulk phase diagrams and bulk nucleation energies, all of which can be straightforwardly computed using DFT. We have applied this concept to two chemically and structurally very different materials systems - intermetallic Fe-Nb alloys and Mg alloyed with small concentrations of Al and Ca - and have demonstrated that MDPDs accurately predict the atomic structure and process conditions for their formation. In both cases, the presence of excess solutes resulted in the formation of planar defects that are confined in thickness and represent local energy minima. Consequently, these defect structures form long-lived metastable defect phases and cannot serve as nuclei for the thermodynamically stable bulk precipitates because further growth would increase their energy. Therefore, the presence of these defect states does not reduce the nucleation barrier of the bulk precipitate phases. In contrast, the formation of these defects substantially lowers the solute chemical potential, effectively reducing or even eliminating the chemical driving force required for the formation of the bulk precipitates.

In addition to the theoretical framework, our proposed concept also has practical implications and provides a direct link to experimental observations. Specifically, the initial and final chemical potentials of the supersaturated host matrix can be determined based on the net-alloy composition and the alloy composition in defect-free regions of the host, respectively. These compositions can be easily obtained through techniques such as electron backscatter diffraction (EBSD) or atom probe tomography (APT) measurements, while the structure and solute excess can be measured using HR-TEM or APT. It is important to note that the concept is not limited to the two prototype material systems studied here and can be readily applied to other systems as well. This opens up a new avenue for materials designers to control microstructure and properties through partitioning into defects, in the same way that precipitate formation is commonly used. By providing materials scientists with an effective tool to exploit these opportunities, MDPDs offer exciting new possibilities for the design of advanced materials.

\section*{Acknowledgement}
The authors acknowledge financial support by the Deutsche Forschungsgemeinschaft (DFG) through the
projects  A06,  B01, and C05 of the CRC1394 ``Structural and Chemical Atomic Complexity -- From Defect Phase Diagrams to Material Properties'', project ID 409476157.

\bibliographystyle{naturemag}
\bibliography{bibliography.bib}
\end{document}